\PassOptionsToPackage{table}{xcolor}
\PassOptionsToPackage{numbers,sort&compress}{natbib}
\documentclass{article}
\usepackage[preprint]{neurips_2026}

\usepackage[utf8]{inputenc}
\usepackage[T1]{fontenc}
\usepackage{hyperref}
\usepackage{url}
\usepackage{booktabs}
\usepackage{amsmath}
\usepackage{amssymb}
\usepackage{amsthm}
\usepackage{amsfonts}
\usepackage{nicefrac}
\usepackage{microtype}
\usepackage{xcolor}
\usepackage{graphicx}
\usepackage{subcaption}
\usepackage{enumitem}
\usepackage{multirow}
\usepackage{array}
\usepackage{mathtools}
\usepackage{algorithm}
\usepackage{algpseudocode}
\usepackage[capitalize]{cleveref}
\crefname{figure}{Fig.}{Figs.}\Crefname{figure}{Fig.}{Figs.}
\crefname{table}{Tab.}{Tabs.}\Crefname{table}{Tab.}{Tabs.}
\crefname{section}{Sec.}{Secs.}\Crefname{section}{Sec.}{Secs.}
\crefname{subsection}{Sec.}{Secs.}\Crefname{subsection}{Sec.}{Secs.}
\crefname{subsubsection}{Sec.}{Secs.}\Crefname{subsubsection}{Sec.}{Secs.}
\crefname{algorithm}{Alg.}{Algs.}\Crefname{algorithm}{Alg.}{Algs.}
\crefname{equation}{Eq.}{Eqs.}\Crefname{equation}{Eq.}{Eqs.}

\theoremstyle{plain}
\newtheorem{theorem}{Theorem}
\newtheorem{proposition}{Proposition}
\newtheorem{lemma}{Lemma}
\theoremstyle{definition}
\newtheorem{definition}{Definition}
\newtheorem{assumption}{Assumption}

\crefname{assumption}{Asm.}{Asms.}\Crefname{assumption}{Asm.}{Asms.}
\crefname{theorem}{Thm.}{Thms.}\Crefname{theorem}{Thm.}{Thms.}
\crefname{proposition}{Prop.}{Props.}\Crefname{proposition}{Prop.}{Props.}
\crefname{lemma}{Lem.}{Lems.}\Crefname{lemma}{Lem.}{Lems.}
\crefname{definition}{Def.}{Defs.}\Crefname{definition}{Def.}{Defs.}
\DeclareMathOperator{\AUROC}{AUROC}
\newcommand{\Pfam}{P_{\mathrm{fam}}}
\newcommand{\Pnew}{P_{\mathrm{new}}}
\newcommand{\KS}{\mathrm{KS}}

\newcommand{\method}{\textsc{CCD}}
\newcommand{\methodv}{\textsc{CCD-AW}}
\definecolor{rankA}{HTML}{FDE047}\definecolor{rankB}{HTML}{FEF08A}\definecolor{rankC}{HTML}{FEF9C3}
\newcommand{\first}[1]{\cellcolor{rankA}\textbf{#1}}
\newcommand{\second}[1]{\cellcolor{rankB}\underline{#1}}
\newcommand{\third}[1]{\cellcolor{rankC}\textit{#1}}

\title{CCD: Context-Calibrated Open-Entity Out-of-Distribution Detection in In-Context Tabular Foundation Models}

\author{%
  Zhangkai Wu\\
  The University of Sydney\\
  \texttt{zhangkai.wu@sydney.edu.au}
  \And
  Xu Zhang\\
  Macquarie University\\
  \texttt{xu.zhang12@hdr.mq.edu.au}
  \AND
  Zhihao Hao\\
  Beijing Technology and Business University\\
  \texttt{hao.zhihao@connect.um.edu.mo}
  \And
  Zhihong Cui\\
  University of Oslo\\
  \texttt{zhihongc@ifi.uio.no}
  \AND
  Yongmin Yoo\\
  Macquarie University\\
  \texttt{yongmin.yoo@mq.edu.au}
  \And
  Xianxun Zhu\\
  Macquarie University\\
  \texttt{xianxun.zhu@mq.edu.au}
  \AND
  Kaize Shi\\
  University of Southern Queensland\\
  \texttt{Kaize.Shi@unisq.edu.au}
}

\begin{document}
\maketitle

\begin{abstract}
In-context tabular foundation models (TFMs), such as TabPFN, predict from a labelled context without updating their parameters and are increasingly
used for out-of-distribution (OOD) detection in Web applications, whose test rows come from entities absent from that
context. Such applications need open-entity OOD detection: a new entity must not be mistaken for an unknown class, and
an unknown class must still be caught. Existing detectors fail in this setting, and an \emph{entity-split gap} hides the failure: random splits also place the
test entities in the context. The failure arises because every TFM score reflects how unfamiliar a row is to the context. This produces an \emph{entity-over-class bias}: known-class rows of new entities look more OOD than familiar ones, sometimes more than unknown classes. Specifically, the two OOD scores most widely used with in-context TFMs face an \emph{entity--class trade-off}: the TFM's predictive confidence misses unknown
classes, whereas the distance of a test row to the context flags new entities. To address this problem, we hold a group of entities out of the context and score them with
the remaining context, which turns them into new entities of known classes and tells us how such rows should score. To measure detection on new entities, we build NEO-Tab (New-Entity OOD detection on
Tables), 14 grouped and temporal tables on which all test rows come from new entities. We then propose
Context-Calibrated Detection (CCD), which uses these group-out swaps on in-distribution data to calibrate
three distances and the confidence against new-entity references. CCD weights them by their sensitivity and sharpness,
keeps the model frozen and uses no OOD data. On NEO-Tab, CCD outperforms 16 existing detectors on every metric we report, and its lead holds on three TFM backbones.
\end{abstract}

\section{Introduction}
\label{sec:intro}

\begin{figure}[tbp]
  \centering
  \includegraphics[width=0.72\linewidth]{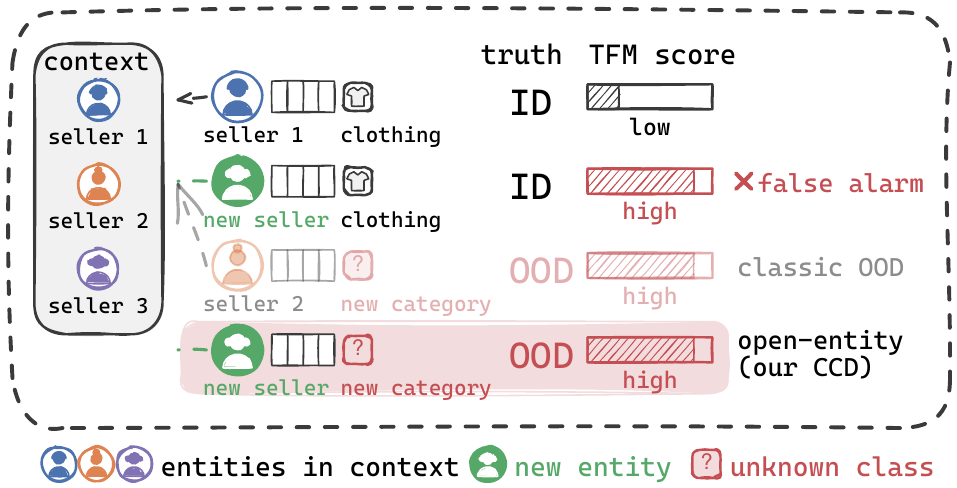}
  \caption{Open-entity OOD detection in a marketplace, with sellers as entities and product categories as classes.
  A high score of a TFM-based detector indicates OOD. In classic OOD detection (faded row), it can only signal an unknown class
  because the entity is in the context. In open-entity OOD detection (highlighted row), known-class (ID) and unknown-class (OOD) listings
  both receive a high score because test rows come from new entities. The detector must separate them, yet existing
  detectors raise a false alarm on the ID listing.}
  \label{fig:problem}
\end{figure}

\begin{figure}[tbp]
  \centering
  \includegraphics[width=\linewidth]{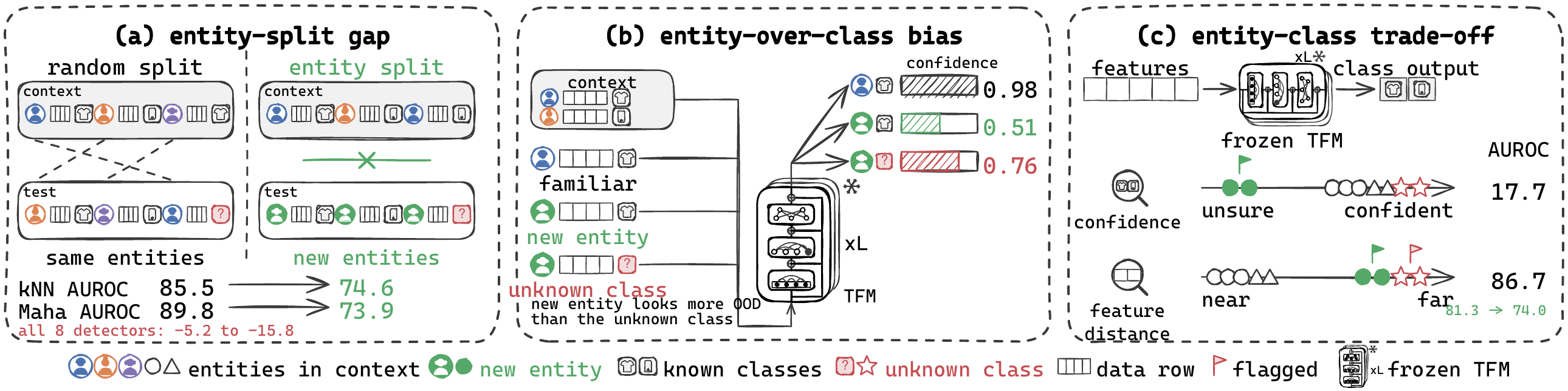}
  \caption[Why open-entity OOD detection is hard for in-context TFMs]{Why open-entity OOD detection is hard for in-context TFMs. Symbols as in \cref{fig:problem}.
  (a) \emph{Entity-split gap}: entity splits cost all eight detectors 5.2--15.8 AUROC points relative to random splits.
  (b) \emph{Entity-over-class bias}: on gesture phase, new-entity ID rows look more OOD than unknown-class rows under confidence. TabPFN v2 confidence there is 0.51 on new-entity rows, below the 0.76 on unknown-class rows and the 0.98 on familiar rows.
  (c) \emph{Entity--class trade-off}:
  feature distance detects the unknown class (AUROC 86.7 against 17.7 for confidence) but also flags new entities.}
  \label{fig:motivation}
\end{figure}

Out-of-distribution (OOD) detection \citep{yang2024generalized} with in-context tabular foundation models (TFMs), which predict from a labelled context without updating their parameters \citep{muller2022pfn,hollmann2025tabpfn,ma2025tabdpt}, is increasingly
deployed in Web services. These services predict for entities, such as merchants in fraud detection \citep{shen2026fraud}
or users on recommendation platforms \citep{wang2022oodrec,sayyadiharikandeh2020novel}, so their test rows come from
entities absent from the context. The detector must therefore reject rows of unknown classes while accepting the
known-class rows of new entities, a problem we call \textbf{open-entity OOD detection}. \Cref{fig:problem} illustrates
it on a marketplace that classifies listings, where a listing from a new seller in a known product category must be
accepted, whereas a listing in a category absent from the context must be rejected.

Open-entity OOD detection first faces an \textbf{entity-split gap}, illustrated in \cref{fig:motivation}a, because random splits conceal how existing detectors fail on
new entities. Tabular OOD detection draws on two kinds of evidence,
confidence and distance, each resting on a premise that a new entity breaks. Confidence-based scores such as the maximum
softmax probability (MSP) \citep{hendrycks2017msp} and its variants \citep{hendrycks2022mls,liu2020energy} assume
that in-distribution (ID) rows are predicted confidently. Distance-based scores such as the Mahalanobis distance
\citep{lee2018maha} and $k$-nearest neighbours (kNN) \citep{sun2022knn} assume that ID rows lie near the training
data. Both premises fail on a known-class row of a new entity, because it is neither predicted confidently nor close
to the context. Random splits \citep{cheng2025realistic} hide this failure because they place the entities of the ID test
rows in the context as well, where both premises hold. As a result,
replacing only the ID test rows of familiar entities with those of new entities lowers the area under the ROC curve
(AUROC) of all eight detectors by 5--16 points on six tasks and changes their ranking.

Open-entity OOD detection also shows an \textbf{entity-over-class bias}: confidence-based and distance-based detectors rank known-class rows of new entities as more OOD than those of familiar entities. With one fixed context, this ranking holds for 13 of the 14 detectors. \Cref{fig:motivation}b shows a case in which these rows look more OOD than even unknown-class rows. On gesture phase, TabPFN with the same
context has a median confidence of 0.98 on known-class rows of familiar entities. On known-class rows of new entities
it drops to 0.51, below even the 0.76 it gives the unknown class. A likely cause lies in in-context prediction, because every score of an in-context TFM is determined by the context. In a simple model of in-context prediction, the confidence
measures how close a test row is to the context and reverts to the prior far from it whatever the class, as
\cref{prop:familiarity} states. Distance-based scores likewise measure unfamiliarity to the context by construction. To an in-context TFM, a new entity and an unknown class can therefore look equally unfamiliar. Yet a known-class row of a new entity should be accepted because the entity changes without changing the class, a covariate shift. An unknown-class row should be rejected because its class is absent from the context, a semantic shift.

The two kinds of evidence also fail in different places, which creates an \textbf{entity--class trade-off}: distances react to both kinds of novelty and see
the unknown class that confidence misses but are themselves sensitive to new entities, as \cref{fig:motivation}c shows.
Confidence fails because a label-oriented output need not retain the features in which an unknown class differs. With TabPFN, MSP stays below 24 AUROC on four tables, while a feature-space kNN reaches 84--87 on three of
them. Distances fail because they carry no class information and therefore flag a new entity far from the context as an
outlier. These complementary failures motivate combining the two kinds of evidence after calibrating each against a reference for how
known-class rows of new entities score.

The property that makes new entities look unfamiliar also lets us create new entities from the context, and our
solution builds on it. To an in-context TFM, a new entity is simply an entity absent from the context, and which entities are
absent is under our control. Holding a group of entities out of the context and scoring them with the remaining
context turns them into new entities of known classes, and under exchangeability their scores are samples from the
new-entity distribution, as \cref{prop:swap} states. Against the entity-split gap, we build NEO-Tab (\textbf{N}ew-\textbf{E}ntity
\textbf{O}OD detection on \textbf{Tab}les), 14 grouped and temporal tables that include the six qualifying BeyondArena tasks
\citep{purucker2026beyond}. On NEO-Tab, ID and OOD test rows both come only from new entities. Against the entity-over-class bias and the entity--class trade-off, we propose
\textbf{C}ontext-\textbf{C}alibrated \textbf{D}etection (CCD), whose group-out context swaps give, from ID data alone, the reference
distribution of every score on new entities and its sensitivity to them. The same swaps give the fraction of held-out
rows on which the confidence stays sharp. CCD corrects the entity-over-class bias by calibrating the confidence and three
distances against their reference distributions, with one distance in a label subspace. \method{} addresses the entity--class trade-off by weighting each distance by
its sensitivity and the confidence by its sharpness before summing them into one score, as \cref{fig:method} shows.

\method{} is the strongest of the 17 detectors on NEO-Tab on all three metrics, gains 7.4 AUROC points over a kNN tuned
to be least sensitive to new entities ($p=0.0012$), and keeps this lead on three backbones. The three testable
predictions of the theory hold on the development trials.

We summarise our contributions as follows.
\begin{itemize}[leftmargin=1.2em,topsep=2pt,itemsep=1pt]
\item We give the first formulation of open-entity OOD detection for in-context tabular models and trace its
entity-split gap, entity-over-class bias and entity--class trade-off to in-context prediction.
\item We build NEO-Tab, 14 grouped and temporal tables on which all ID and OOD test rows come from unseen entities, and
release its splits and scripts.
\item We propose \method{}, whose group-out swaps give a new-entity reference from ID data alone, against which four
scores are calibrated and weighted before fusion.
\item We find \method{} to be the strongest of 17 detectors on 14 tables, 7.4 AUROC points above a kNN tuned to be least
sensitive to new entities, with a lead that holds on three TFM backbones. The three testable predictions of our theory hold on the development trials.
\end{itemize}

\section{Related Work}
\label{sec:related}

\textbf{Tabular foundation models under shift.} In-context TFMs such as TabPFN-2.5 \citep{grinsztajn2025tabpfn25} and
TabICL \citep{qu2025tabicl} now lead tabular benchmarks, and Web applications already use them for graph anomaly detection
\citep{liu2026tfmgraph} and evolving data streams \citep{lourenco2026streams}. Work on these models under shift asks how
accurate prediction stays on new entities or later periods, not whether a row belongs to a known class. StablePFN
\citep{guan2026stablepfn}, Drift-Resilient TabPFN \citep{helli2024drift} and a mixture of in-context experts for recurring concept drift \citep{wu2026recall} make prediction robust to these shifts, and
grouped benchmarks such as TableShift \citep{gardner2023tableshift} and BeyondArena \citep{purucker2026beyond} measure
how much accuracy they cost. None of these works compares how much more an in-context model loses on new entities than
a trained model does, which is the comparison of our entity-over-class analysis in \cref{sec:failure}: three TFM
backbones lose 8.5--9.2 points of MSP AUROC on new entities, whereas CatBoost and TabM lose 5.1 and 2.1.

\textbf{OOD detection and its evaluation.} Existing OOD benchmarks share entities between training and test, so they
never ask a detector to accept a new entity while rejecting an unknown class. OpenOOD \citep{yang2022openood,
zhang2024openoodv15} evaluates confidence-based and distance-based detectors against a fixed training set, and methods
based on outlier exposure \citep{hendrycks2019oe}, outlier synthesis \citep{du2022vos} or weakly augmented variational autoencoders \citep{wu2026wavae} additionally train the network,
which a frozen in-context model does not allow. Full-spectrum detection \citep{yang2023fullspectrum} accepts covariate
shift and rejects semantic shift, but it defines both shifts relative to a fixed training set, not to a context that a
frozen model reads at inference. Tabular evaluations either use random splits \citep{cheng2025realistic,han2022adbench}
or treat a new patient group as OOD \citep{ulmer2020trust,azizmalayeri2025unmasking}. NEO-Tab in \cref{sec:benchmark}
measures the setting that these evaluations leave out.

\textbf{Calibration with held-out data.} Mondrian conformal prediction \citep{vovk2012conditional}
and leave-group-out cross-validation \citep{roberts2017crossvalidation} also hold groups out, to calibrate prediction
sets per group or to estimate performance on unseen groups, but both refit the model or need a held-out calibration set.
Our context swaps instead calibrate OOD scores for new entities and need nothing beyond forward passes of a frozen
model.

\section{Open-Entity OOD Detection}
\label{sec:observation}

\subsection{Problem Definition}
\label{sec:problemdef}

Rows $(\mathbf{x},y,g)$ carry an entity $g$ such as a user or a region. A time block also counts as an entity,
and we call the entity of a row its \emph{group}.
Given a context $C$, let $\Pfam$ and $\Pnew$ be the laws of ID rows whose entity does or does not occur in $C$, and
$Q_{\mathrm{fam}}$, $Q_{\mathrm{new}}$ those of OOD rows. For a score $s$ (high means OOD),
$\AUROC(s;P,Q)=\Pr(s(\mathbf{Z})>s(\mathbf{X}))$ with $\mathbf{Z}\sim Q$, $\mathbf{X}\sim P$ independent; we assume
continuous scores, write $G_P$ for the CDF of $s(\mathbf{X})$, $\mathbf{X}\sim P$, and
$\KS(P,P')=\sup_t|G_P(t)-G_{P'}(t)|$.

\begin{definition}[Open-entity OOD detection]\label{def:openentity}
An in-context detector scores each test row given a labelled context $C$. In \emph{open-entity OOD detection}, ID
rows follow $\Pnew$ and rows of an unknown class follow $Q_{\mathrm{new}}$; the detector must accept the former, a
covariate shift, and reject the latter, a semantic shift. Its target is
$A_{\mathrm{new},\mathrm{new}}=\AUROC(s;\Pnew,Q_{\mathrm{new}})$.
\end{definition}

How differently a detector treats new and familiar ID entities is measured by the sensitivity
$S(s)=\Pr(s(\mathbf{X}')>s(\mathbf{X}))+\frac12\Pr(s(\mathbf{X}')=s(\mathbf{X}))$ with $\mathbf{X}'\sim\Pnew$,
$\mathbf{X}\sim\Pfam$: $S=\frac12$ means that new entities are not systematically ranked as more or less OOD, and we report
$S$ in percent and $|S-50|$.

\begin{lemma}[Evaluations as mixtures]\label[lemma]{lem:mixture}
If the ID test rows follow $P_\pi=(1-\pi)\Pfam+\pi\Pnew$ and the OOD test rows follow
$Q_\tau=(1-\tau)Q_{\mathrm{fam}}+\tau Q_{\mathrm{new}}$, then $\AUROC(s;P_\pi,Q_\tau)$ is the bilinear
interpolation of the four AUROCs $A_{ab}=\AUROC(s;P_a,Q_b)$, $a,b\in\{\mathrm{fam},\mathrm{new}\}$, with weights
$(1-\pi,\pi)$ and $(1-\tau,\tau)$.
\end{lemma}

Open-entity OOD detection is the corner $\pi=\tau=1$.

\subsection{Why Random-Split Evaluation Misses the Problem}
\label{sec:pitfalls}\label{sec:benchmark}

Random-split evaluation of tabular OOD detection \citep{cheng2025realistic} measures another
point of this square, for three reasons. \textbf{P1: familiar ID rows} ($\pi\approx0$): a detector that ranks new entities as more
OOD is overestimated; a bound limits the gap by the Kolmogorov--Smirnov distance between
familiar and new entities and shows that it requires $S\neq50$. \textbf{P2: familiar OOD rows} ($\tau<1$): on the six BeyondArena
tasks 62--92\% of the OOD rows come from entities of the training split, so the ID--OOD contrast mixes the class shift
with the entity shift. \textbf{P3: batch-normalised scores}: a good detector looks tolerant however far new entities move,
because ranking scores within the test batch keeps only the OOD share of the shift.

Our protocol and the 14 tables of \cref{sec:setup} form the NEO-Tab benchmark. The protocol measures $A_{\mathrm{new},\mathrm{new}}$ directly: ID test rows come from the official test split (new entities or future), OOD rows only from entities absent from the training split or from after its end, and no
score is normalised within the batch. Each trial is paired with a random-split trial of the same size on the same
OOD rows, and the two therefore differ only in the ID rows.

\section{Context-Calibrated Detection}
\label{sec:method}
\begin{figure}[tbp]
  \centering
  \includegraphics[width=\linewidth]{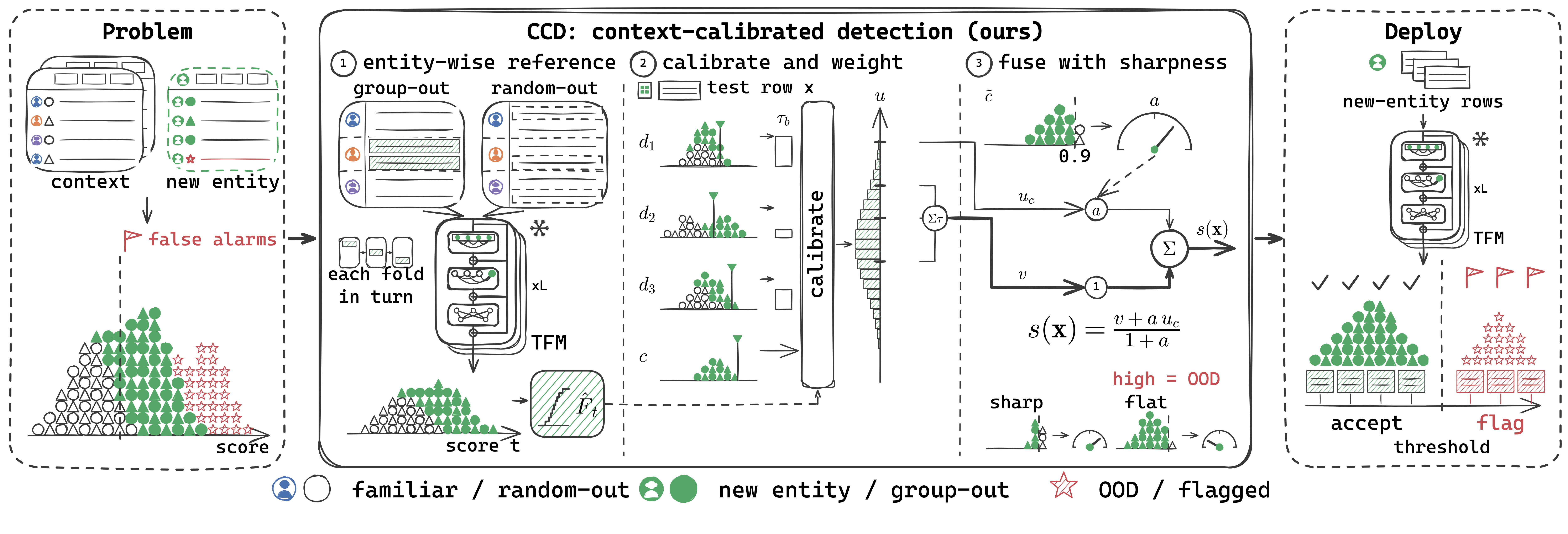}
  \caption{The \method{} framework: (1) group-out swaps give a new-entity reference, (2) distances and confidence are
  calibrated against it and weighted, and (3) the confidence is added with sharpness $a$.}
  \label{fig:method}
\end{figure}

NEO-Tab addresses the entity-split gap in evaluation, and a detector evaluated on it needs a reference for how known-class rows of new entities score. The context can supply one because to an in-context TFM a new entity is simply one absent from the context. Holding a group of entities out of the context and scoring its rows
with the remaining context therefore yields known-class rows of new entities from ID data alone. Module (1) of \cref{fig:method}, Steps 1--2 of \cref{sec:ccdscore}, turns these rows into a reference distribution for every score.

The entity-over-class bias requires the confidence and each distance alike to be calibrated against this new-entity
reference, not against familiar rows. Calibration against familiar rows fails because far from the context the
confidence reverts to the prior whatever the class, as \cref{prop:familiarity} states, and a distance treats distance from the context as outlyingness by construction. The same swaps measure how systematically a
distance ranks new entities above or below familiar ones, and this directional sensitivity sets its weight among the distances in module (2), Steps 3--4.

The entity--class trade-off rules out dropping either kind of evidence, because confidence and distances fail in different
places. A label-oriented output can discard the features in which an unknown class differs, whereas distances also respond to new entities. The distance branch therefore keeps
weight one, and the confidence is added with a weight equal to how sharp it stays on held-out entities in module (3), Steps 5--6. \Cref{sec:theory} states what each module guarantees.

\subsection{The CCD Score}
\label{sec:ccdscore}

CCD uses the labelled context itself to approximate how known-class rows behave when their entity is absent, and
calibrates complementary detection evidence against this reference. It combines a label branch, the confidence,
with a distribution branch of distances. We instantiate both for a frozen in-context learner $f_{\Theta}(\cdot\,;C)$ with context
$C=\{(\mathbf{x}_i,y_i,g_i)\}_{i=1}^{n}$, where $g_i$ is the group (or time block) of row $i$, and calibrate each branch on
the reference of \cref{sec:pitfalls}, ID rows of \emph{new} groups, as \cref{fig:method} shows.

\textbf{Step 1: group-out and random-out context swaps.} Partition $C$ by group into folds $F_1,\dots,F_m$ ($m\le5$; temporal tasks use
five time blocks). For a row score $t(\cdot\,;C)$ and $i\in F_l$, the \emph{group-out} value is $\tilde
t_i=t(\mathbf{x}_i;C\setminus F_l)$; the \emph{random-out} value $\bar t_i$ uses five random folds instead. Both cost
forward passes only, because $C\setminus F_l$ is a new context for a TFM. The group-out values are a sample of the score
under $\Pnew$ and the random-out values a sample under $\Pfam$ by \cref{prop:swap}, and both use ID rows alone.
\textbf{Step 2: new-entity reference distributions.} The swaps estimate the new-group null of every score, $\hat F_t(\omega)=\bigl(\sum_i\mathbf{1}[\tilde t_i\le \omega]+\frac12\bigr)/(n+1)$ for $\omega\in\mathbb{R}$ (the empirical CDF, smoothed to keep normal scores finite), with normal score
$u_t(\mathbf{x})=\Phi^{-1}(\hat F_t(t(\mathbf{x})))$. Comparing group-out with random-out values gives the covariate-shift sensitivity of \cref{sec:problemdef},
$\hat S_t=\widehat{\Pr}(\tilde t_i>\bar t_j)$, without access to any test row.

\textbf{Step 3: distribution branch.} This branch uses three distances to the context, each standardised on $C$.
The feature-space distance $d_1$ is the mean distance to the $k=10$ nearest context rows, and $d_2$ is the same distance in the
\emph{label subspace}. This subspace spans the at most $K-1$ discriminant directions $\mathcal{W}=\Sigma_W^{-1}[\mathbf{m}_y-\bar{\mathbf{m}}]_y$ of the context
classes, with a shrinkage estimate of the within-class covariance $\Sigma_W$. The embedding distance $d_3$ is the Mahalanobis distance in the
TFM embedding. \textbf{Step 4: calibration and shift-dependent weighting.} In normal scores under their new-group
nulls, the three distances are combined with \emph{tolerance} weights that favour the distances least sensitive to new groups,
\begin{equation}
  v(\mathbf{x})=\frac{1}{\hat\sigma}\sum_{b=1}^{3}\tau_b\,u_{d_b}(\mathbf{x}),\qquad \tau_b\propto 1-2\bigl|\hat S_{d_b}-\tfrac12\bigr|,
  \label{eq:dist}
\end{equation}
where $\hat\sigma$, the standard deviation of $\sum_b\tau_b\tilde u_{d_b}$ over the group-out values with
$\tilde u_{d_b}=\Phi^{-1}((\mathrm{rank}(\tilde d_{bi})-\frac12)/n)$, scales $v$ to unit variance under the new-group null. The weights $\tau_b$ down-weight a distance that ranks
new groups systematically higher or lower than familiar ones. A distance that is equally sensitive in both
directions keeps its weight, because such shifts widen its null without moving the reference.

\textbf{Step 5: label branch and sharpness.} $c(\mathbf{x})=\max_y p_{\Theta}(y\mid \mathbf{x};C)$ with normal score $u_c(\mathbf{x})=\Phi^{-1}(\hat F_{-c}(-c(\mathbf{x})))$. Its
weight is the \emph{sharpness} of the confidence on new groups,
\begin{equation}
  a=\frac1n\sum_{i=1}^{n}\mathbf{1}[\tilde c_i>\theta],\qquad\theta=0.9,
  \label{eq:sharp}
\end{equation}
the share of context rows that the TFM still predicts confidently when their own group is removed.

\textbf{Step 6: CCD score.} The two branches are combined additively,
\begin{equation}
  s(\mathbf{x})=\frac{v(\mathbf{x})+a\,u_c(\mathbf{x})}{1+a},
  \label{eq:score}
\end{equation}
where a high $s$ indicates OOD. In other words, the distance keeps weight one and the confidence is \emph{added} with weight $a\in[0,1]$.
The score $s$ is a fixed function of $\mathbf{x}$ given $C$: no quantity is normalised within the test batch, following P3
of \cref{sec:pitfalls}. In summary, the swaps define the new-entity reference, measure the sensitivity $\hat S$ that sets the weights,
and estimate the sharpness $a$. \Cref{alg:ccd} summarises the procedure.

\begin{algorithm}[t]
\caption{CCD: Context-Calibrated Detection}
\label{alg:ccd}
\small
\begin{algorithmic}[1]
\Require frozen in-context TFM $f_\Theta$; context $C=\{(\mathbf{x}_i,y_i,g_i)\}_{i=1}^n$; folds $F_1,\dots,F_m$ by group; $\theta=0.9$, $k=10$
\Ensure OOD score $s(\mathbf{x})$ for every test row $\mathbf{x}$ (high means OOD)
\Procedure{Calibrate}{$C$} \Comment{once per context, forward passes only}
  \State $\mathcal{W}\gets\Sigma_W^{-1}[\mathbf{m}_y-\bar{\mathbf{m}}]_y$ on $C$ \Comment{label subspace of $d_2$}
  \For{$l=1,\dots,m$} \Comment{group-out swaps, Step 1}
    \For{$i\in F_l$ and $t\in\{-c,d_1,d_2,d_3\}$}
      \State $\tilde t_i\gets t(\mathbf{x}_i;C\setminus F_l)$ \Comment{sample under $\Pnew$, \cref{prop:swap}}
    \EndFor
  \EndFor
  \State $\bar t_i\gets t(\mathbf{x}_i;C\setminus R_l)$, $R_l\ni i$ random \Comment{random-out, under $\Pfam$}
  \State $\hat F_t(\omega)\gets\bigl(\sum_i\mathbf{1}[\tilde t_i\le\omega]+\frac12\bigr)/(n+1)$ \Comment{new-entity null, Step 2}
  \State $\hat S_{d_b}\gets\widehat{\Pr}(\tilde d_{bi}>\bar d_{bj})$ for $b=1,2,3$ \Comment{sensitivity}
  \State $\tau_b\propto\max\bigl(0,1-2|\hat S_{d_b}-\tfrac12|\bigr)$, $\sum_b\tau_b=1$ \Comment{tolerance weights, \cref{eq:dist}}
  \State $\hat\sigma\gets\operatorname{sd}_i\bigl(\sum_b\tau_b\,\Phi^{-1}((\mathrm{rank}(\tilde d_{bi})-\frac12)/n)\bigr)$
  \State $a\gets\frac1n\sum_i\mathbf{1}[\tilde c_i>\theta]$ \Comment{sharpness, \cref{eq:sharp}}
\EndProcedure
\Procedure{Score}{$\mathbf{x}$} \Comment{fixed function of $\mathbf{x}$ given $C$}
  \State $c\gets\max_y p_\Theta(y\mid\mathbf{x};C)$; $d_1,d_2,d_3\gets$ distances of $\mathbf{x}$ to $C$ \Comment{Step 3}
  \State $u_{d_b}\gets\Phi^{-1}(\hat F_{d_b}(d_b))$, $u_c\gets\Phi^{-1}(\hat F_{-c}(-c))$ \Comment{calibration, Step 4}
  \State $v\gets\hat\sigma^{-1}\sum_{b=1}^{3}\tau_b\,u_{d_b}$ \Comment{distribution branch, \cref{eq:dist}}
  \State \Return $s(\mathbf{x})=(v+a\,u_c)/(1+a)$ \Comment{additive fusion, \cref{eq:score}}
\EndProcedure
\end{algorithmic}
\end{algorithm}

\begin{table}[tbp]
  \centering
  \caption{Detection on the 14 benchmark tables (TabPFN v2, held-out trials 35--39). Rank: average rank among 17
  detectors; Worst: lowest per-table AUROC; Random: random-split protocol; 40--44: resampling check; Best: tables won;
  last two columns: \method{} minus the detector in AUROC with tables won, and Holm-corrected Wilcoxon $p$. The top three
  values per column are shaded.}
  \label{tab:main}
  \small
  \setlength{\tabcolsep}{2pt}
\resizebox{\linewidth}{!}{%
\begin{tabular}{clcccccccccccc}
\toprule[1pt]
 & & \multicolumn{2}{c}{\textbf{AUROC $\uparrow$}} & \multicolumn{2}{c}{\textbf{AUPR $\uparrow$}} & \multicolumn{2}{c}{\textbf{FPR95 $\downarrow$}} & \multicolumn{3}{c}{\textbf{AUROC $\uparrow$}} & \textbf{Best $\uparrow$} & \multicolumn{2}{c}{\textbf{\method{} $-$ detector (AUROC)}} \\
\cmidrule(lr){3-4}\cmidrule(lr){5-6}\cmidrule(lr){7-8}\cmidrule(lr){9-11}\cmidrule(lr){13-14}
& \textbf{Detector} & mean & rank & mean & rank & mean & rank & Worst & Random & 40--44 & tables & $\Delta$ (won) & Holm $p$ \\
\midrule
\multirow{3}{*}{\rotatebox[origin=c]{90}{\footnotesize\itshape \shortstack{TFM\\output}}} & MSP~\citep{hendrycks2017msp} & 56.6 & 10.9 & 32.6 & 11.9 & 68.1 & 10.1 & 13.1 & 67.6 & 55.9 & \third{2} & $+25.2$ (11/14) & 0.037 \\
 & MaxLogit~\citep{hendrycks2022mls} & 57.6 & 10.3 & 33.7 & 10.6 & 68.6 & 9.7 & 12.9 & 67.5 & 55.8 & 0 & $+24.2$ (11/14) & 0.037 \\
 & Energy~\citep{liu2020energy} & 58.0 & 9.6 & 34.5 & 10.3 & 68.4 & 9.0 & 12.6 & 67.7 & 56.2 & 1 & $+23.8$ (11/14) & 0.037 \\
\midrule
\multirow{9}{*}{\rotatebox[origin=c]{90}{\footnotesize\itshape Model-free}} & kNN$^\dagger$~\citep{sun2022knn} & 74.4 & 8.4 & 43.5 & 8.2 & 64.8 & 10.3 & 47.8 & 75.2 & 75.1 & 0 & $+7.4$ (12/14) & 0.012 \\
 & kNN~\citep{sun2022knn} & 74.2 & 7.5 & 43.2 & 8.0 & 59.1 & 7.4 & \third{48.5} & 78.6 & 74.9 & 1 & $+7.6$ (11/14) & 0.032 \\
 & Maha (x)~\citep{lee2018maha} & 71.2 & 9.5 & 40.7 & 9.9 & 63.9 & 8.9 & 46.5 & 73.8 & 71.5 & 1 & $+10.6$ (12/14) & 0.037 \\
 & Maha (emb.)~\citep{lee2018maha} & 73.6 & \third{7.0} & \third{44.8} & \second{6.1} & 59.9 & 7.6 & \first{50.6} & \second{85.6} & 71.8 & 0 & $+8.2$ (10/14) & 0.059 \\
 & Emb-kNN~\citep{sun2022knn} & 67.5 & 8.4 & 43.5 & 7.7 & 64.2 & 8.9 & 18.7 & \third{83.8} & 66.7 & 0 & $+14.2$ (12/14) & 0.037 \\
 & IForest~\citep{liu2008iforest} & 70.5 & 10.1 & 39.7 & 10.3 & 65.4 & 9.9 & 41.4 & 71.3 & 71.0 & 1 & $+11.3$ (11/14) & 0.021 \\
 & LOF~\citep{breunig2000lof} & 66.5 & 10.2 & 38.4 & 10.0 & 70.2 & 10.9 & 31.3 & 74.0 & 67.6 & 0 & $+15.3$ (12/14) & 0.008 \\
 & OC-SVM~\citep{scholkopf2001estimating} & 69.6 & 10.9 & 41.1 & 10.3 & 65.7 & 10.4 & 45.5 & 69.4 & 70.4 & 0 & $+12.2$ (13/14) & 0.005 \\
 & ECOD~\citep{li2022ecod} & 63.1 & 12.4 & 33.1 & 12.4 & 72.1 & 11.9 & 40.2 & 62.5 & 63.6 & 0 & $+18.7$ (14/14) & 0.002 \\
\midrule
\multirow{3}{*}{\rotatebox[origin=c]{90}{\footnotesize\itshape \shortstack{Pre-\\trained}}} & FoMo-0D~\citep{shen2025fomo0d} & 67.6 & 9.7 & 43.1 & 9.4 & 72.4 & 10.7 & 25.4 & 71.0 & 67.0 & 1 & $+14.1$ (13/14) & 0.009 \\
 & OutFormer~\citep{ding2026outformer} & 68.2 & 10.5 & 40.4 & 9.7 & 72.3 & 11.3 & 46.4 & 74.8 & 68.5 & 0 & $+13.5$ (13/14) & 0.004 \\
 & TabPFN-unsup~\citep{hollmann2025tabpfn} & \third{74.9} & \second{6.1} & \second{48.1} & \third{6.4} & \second{54.9} & \second{5.3} & 34.1 & -- & \second{76.0} & \second{3} & $+6.9$ (10/14) & 0.173 \\
\midrule
\multirow{2}{*}{\rotatebox[origin=c]{90}{\footnotesize\itshape Ours}} & \methodv{} & \second{75.1} & 7.8 & 44.0 & 7.6 & \third{58.3} & \third{6.9} & 48.4 & 80.0 & \third{75.8} & 0 & $+6.7$ (12/14) & 0.008 \\
 & \textbf{\method{}} & \first{81.8} & \first{3.6} & \first{52.4} & \first{4.2} & \first{50.3} & \first{3.9} & \second{48.8} & \first{86.5} & \first{81.1} & \first{4} & -- & -- \\
\bottomrule[1pt]
\end{tabular}}
\end{table}

\subsection{Theoretical Guarantees}
\label{sec:theory}

The results below state, for each module, what the context swaps estimate and when the fused score cannot lose to its
distance branch. Proofs are omitted from this preprint.

\textbf{The reference is unbiased.} Under exchangeable groups, the group-out values follow the law of every score on
new entities and the random-out values its law on familiar ones, so the reference of module (1) needs neither OOD data
nor test rows.

\begin{assumption}[Exchangeable groups]\label{ass:groups}
Context groups and test groups are i.i.d.\ draws from $P_G$, rows are i.i.d.\ within a group given $g$, and
$f_{\Theta}(\cdot\,;C)$ and every $d_b$ are invariant to permutations of the context rows.
\end{assumption}

\begin{proposition}[Context swaps sample the new-group distribution]\label[proposition]{prop:swap}
Under \cref{ass:groups}, for $i\in F_l$ the group-out values $(\tilde c_i,\tilde d_{1i},\tilde d_{2i},\tilde d_{3i})$
have the law of $(c(\mathbf{X}'),d_1(\mathbf{X}'),d_2(\mathbf{X}'),d_3(\mathbf{X}'))$ for $\mathbf{X}'\sim\Pnew$, computed with a context of the groups outside
$F_l$; under independent uniform fold assignment, the random-out values follow $\Pfam$ for every row whose group keeps a
row in the reduced context.
\end{proposition}

Hence $\hat F_t$, $a$ and $\hat S_t$ estimate the new-group null, $\Pnew(c>\theta)$ and the sensitivity $S(t)$ of
\cref{sec:problemdef} from the context alone, with blocked time folds treated as groups; $\hat S_t$ is the
Mann--Whitney statistic of the group-out against the random-out values of $t$.

\textbf{Why new entities look unfamiliar.} A simple model explains how the entity-over-class bias can arise. A prior-data fitted network (PFN) approximates the posterior predictive of its prior \citep{muller2022pfn}, and in-context TFMs share its interface of predicting from a labelled context. As an idealisation of this view, we analyse a kernel-weighted in-context predictor. Consider a
Dirichlet prior $\pi$ over $K$ classes with strength $\alpha>0$, updated by kernel-weighted context labels,
\begin{equation}
  p(y\mid \mathbf{x};C)=\frac{\alpha\pi_y+\sum_{i=1}^{n}\beta(\lVert \mathbf{x}-\mathbf{x}_i\rVert)\,\mathbf{1}[y_i=y]}
                   {\alpha+\sum_{i=1}^{n}\beta(\lVert \mathbf{x}-\mathbf{x}_i\rVert)},
  \label{eq:kppd}
\end{equation}
with a non-increasing kernel $\beta:[0,\infty)\to(0,1]$, $\beta(\delta)\to0$ as $\delta\to\infty$. Let
$c_0=\max_y\pi_y$, $c(\mathbf{x})=\max_y p(y\mid\mathbf{x};C)$ and
$\delta(\mathbf{x})=\min_i\lVert \mathbf{x}-\mathbf{x}_i\rVert$.

\begin{proposition}[Confidence as familiarity]\label[proposition]{prop:familiarity}
Under \cref{eq:kppd}, $\alpha c_0/(\alpha+n)\le c(\mathbf{x})\le\psi(\delta(\mathbf{x}))$ with
$\psi(\delta)=c_0+(1-c_0)\,n\beta(\delta)/(\alpha+n\beta(\delta))$, which is non-increasing and tends to $c_0$. With a
uniform prior, if a new ID entity and an OOD class both lie at distance at least $\delta_0$ from the context, their
mean confidences differ by at most $\psi(\delta_0)-1/K$, whatever the class of the rows.
\end{proposition}

Under this model the bias is a property of in-context prediction and not of classifiers in general, because trained ReLU networks
instead grow more confident far from the data \citep{hein2019relu}.

\textbf{What each branch can see.} The label subspace removes \emph{group inflation}, which is the extra feature-space distance that label-irrelevant
entity-level columns add when a row's group is absent. Yet the same projection is blind to classes that differ only there. Likewise, the confidence detects an unknown class only insofar as it stays sharp on
new entities. These two limits are the entity--class trade-off within the branches. Both limits share one cause: a representation trained for the label need
not keep anything else, and unknown classes can then collapse onto known features
\citep{vanamersfoort2020duq}. Every confidence-based score is therefore limited to
what the label representation retains. When the label-irrelevant columns have block-diagonal within-class covariance, the label-subspace distance $d_2$
is invariant to shifts of label-irrelevant columns, which inflate the feature-space distance $d_1$ in proportion to the
number of entity-level columns. Yet $d_2$ has AUROC $\frac12$ against a class that differs only in those columns, and
\cref{eq:dist} therefore keeps both distances. The AUROC of the confidence against an unknown class lies between $a_\theta q_\theta$ and $1-(1-a_\theta)(1-q_\theta)$, where $a_\theta=\Pnew(c(\mathbf{X}')>\theta)$ is the share of
new-entity ID rows still predicted above $\theta$ and $q_\theta$ the share of OOD rows below it. Because an unknown class far from the context receives a confidence near the prior by \cref{prop:familiarity}, the
confidence detects it only to the extent that ID rows of new groups stay sharp, and the entity-over-class bias lowers
$a_\theta$. The weight $a$ of \cref{eq:sharp} estimates $a_\theta$ at $\theta=0.9$ by context swaps.

\textbf{Fusion keeps the distance.} Adding the confidence with the sharpness weight can lower the AUROC of the distance branch
only by the amount bounded in \cref{thm:fusion}(ii), and never discards the distance. In population form, let $U=u_c(\mathbf{x})$ be the normal
score of $-c$ and $V=v(\mathbf{x})$ that of
\cref{eq:dist}; under $\Pnew$ both have mean 0 and variance 1, and we assume joint normality.

\begin{assumption}[Equal-covariance Gaussian model of the two branches]\label{ass:copula}
$(U,V)\sim\mathcal{N}(\mathbf{0},\mathcal{R})$ under $\Pnew$ and $(U,V)\sim\mathcal{N}(\boldsymbol{\mu},\mathcal{R})$ under $Q$, with $\boldsymbol{\mu}=(\mu_U,\mu_V)$,
$\mathcal{R}=\bigl(\begin{smallmatrix}1&r\\ r&1\end{smallmatrix}\bigr)$ and $|r|<1$.
\end{assumption}

The branch AUROCs are $A_c=\Phi(\mu_U/\sqrt2)$ and $A_d=\Phi(\mu_V/\sqrt2)$, and \cref{eq:score} is
$s_w=wU+(1-w)V$ with $w=a/(1+a)\in[0,\frac12]$.

\begin{theorem}[Additive fusion with a sharpness weight]\label{thm:fusion}
Let \cref{ass:copula} hold with $\mu_V\ge0$ and $w=a/(1+a)$, $a\in(0,1]$. Then:
\begin{enumerate}
\item[(i)] (When the confidence helps.) $\AUROC(s_w)\ge A_d$ if and only if $\mu_U\ge\kappa(w,r)\,\mu_V$, where
\[
  \kappa(w,r)=\frac{\sqrt{1-2w(1-w)(1-r)}-(1-w)}{w}
\]
increases in $w$ from $\kappa(0^+,r)=r$ to $\kappa(\frac12,r)=\sqrt{2(1+r)}-1<1$.
\item[(ii)] (Bounded loss.) If $\mu_U\le\mu_V$ and $w\mu_U+(1-w)\mu_V\ge0$, then
$\AUROC(s_w)\ge A_d-\dfrac{a}{1+a}\cdot\dfrac{\mu_V-\mu_U}{2\sqrt\pi}$.
\item[(iii)] (The distance is never discarded.) For $\mu_U\ge0$, $\AUROC(s_w)\ge\Phi\bigl(\mu_V/(2\sqrt2)\bigr)$ for every $a\in[0,1]$, whereas the convex form $(1-a)V+aU$ has AUROC $A_c$ at $a=1$, however large
$\mu_V$.
\end{enumerate}
\end{theorem}

A small weight asks little of the confidence: as $w\to0$, (i) only requires the part of the confidence not explained
by the distance to shift towards OOD, while a sharp confidence ($a=1$) must reach $\sqrt{2(1+r)}-1$ of the distance
shift. When the confidence is blind ($\mu_U<0$), the loss is at most proportional to $a/(1+a)$ (ii), and adding the confidence
instead of mixing it in keeps at least half of the weight on the distance (iii).

\begin{figure}[tbp]
  \centering
  \begin{minipage}[t]{0.48\linewidth}
  \centering
  \includegraphics[width=\linewidth,trim=0 0 262 13,clip]{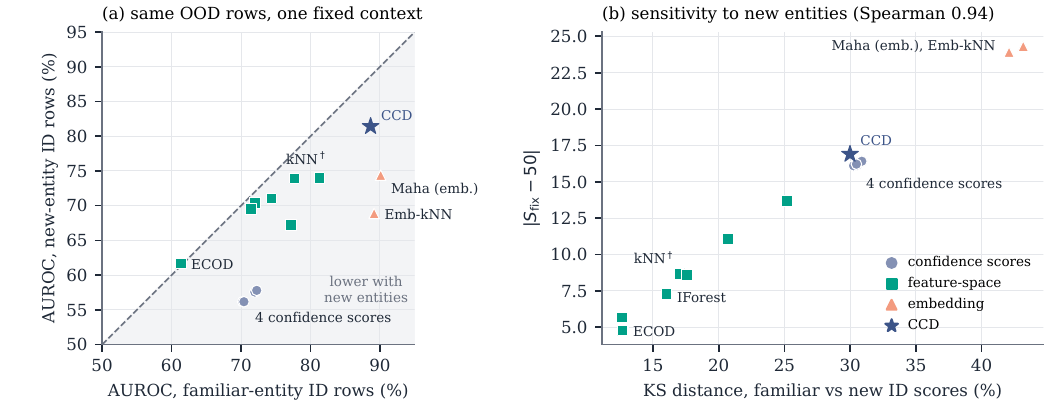}
  \caption{Fixed-context check (TabPFN v2, held-out trials 35--39, 14 tables): AUROC against the same OOD rows with
  familiar-entity and with new-entity ID rows, one context per trial; 13 of 14 detectors fall below the diagonal.}
  \label{fig:app-fixedctx}
  \end{minipage}\hfill
  \begin{minipage}[t]{0.48\linewidth}
  \centering
  \includegraphics[width=\linewidth,trim=252 0 0 13,clip]{figures/app_fixedctx.pdf}
  \caption{Two measures of how far ID scores move from familiar to new entities agree (Spearman 0.94 over the 14
  detectors).}
  \label{fig:app-fixedctx-ks}
  \end{minipage}
\end{figure}

\section{Experiments}
\label{sec:experiments}

We organise the experiments around four research questions:
\begin{itemize}
  \item \textbf{RQ1}: how much does random-split evaluation overestimate detectors?
  \item \textbf{RQ2}: how do the entity-over-class bias and the entity--class trade-off affect existing detectors and backbones?
  \item \textbf{RQ3}: how does \method{} compare with existing detectors on new entities?
  \item \textbf{RQ4}: which components of \method{} contribute?
\end{itemize}

\subsection{Experimental Setup}
\label{sec:setup}

\textbf{Datasets.} We use all six multiclass BeyondArena tasks with a grouped or temporal split whose smallest class
has at least 30 rows, four with grouped and two with temporal splits. We add ten tables whose
rows are grouped by an entity such as a subject or a course, with reserved test groups. All sixteen tables follow five selection rules. R1 requires a multiclass target ($\ge3$ classes),
and R2 a group structure, either an official grouped or temporal split or, in an IID table, a column that names a real
entity and is not the label. R3 keeps a table within the stated limits of TabPFN v2 ($\le500$ features, $\le10$ known
classes), the restriction TabArena applies to TabPFN v2 \citep{erickson2025tabarena}, and R4 requires an OOD class, the
smallest class with at least 30 rows. R5 excludes tables outside BeyondArena that appear in the pretraining collections
of TabDPT and Real-TabPFN-2.5. Applying R1--R4 to the metadata of all 142 BeyondArena tasks \citep{purucker2026beyond}
gives 25 multiclass tasks, of which 8 have an official grouped or temporal split and 2 of the 17 IID ones have an
entity column, the plate in sdss17 and the course in students. After R3 removes micro\_mass (1{,}082 features), all remaining nine
pass R4, and the six whose smallest class itself has at least 30 rows form the six-task study. The remaining asp, whose
smallest class has 21 rows and whose OOD class is therefore the smallest with at least 30, joins sdss17 and students as
an entity table. \textbf{Other public tables.} Seven public tables outside BeyondArena pass R1--R5 on their metadata and are added, and N-BaIoT also passes but is left out for its size of about seven million rows. The
OOD class is the smallest class with at least 30 rows and never enters the context, and all ID and OOD test rows come
from new groups. The backbone is TabPFN v2, and the metrics are AUROC, AUPR and FPR95. The 14 tables that enter the
held-out evaluation after the size rule (at least 10 new-group ID and OOD test rows per trial) form NEO-Tab with the protocol of \cref{sec:benchmark},
and we release row identifiers alongside the splits and scripts for all sixteen tables.

\textbf{Baselines.} We compare against 16 detectors: TFM output scores, model-free feature-space and embedding
detectors, and pretrained outlier detectors, listed in \cref{tab:main}. kNN$^\dagger$ tunes $k$ on ID data with the
same signal as \method{}.

\textbf{Protocol.} \method{} is designed on development trials 0--4 and evaluated on held-out
trials 35--39, which are drawn from reserved test groups disjoint from the development groups. The primary test compares \method{} with kNN$^\dagger$ by a two-sided Wilcoxon signed-rank test over tables. Trials 40--44 redraw rows and seeds from the same tables and reserved groups as a resampling check.

\begin{figure}[tbp]
  \centering
  \includegraphics[width=0.6\linewidth]{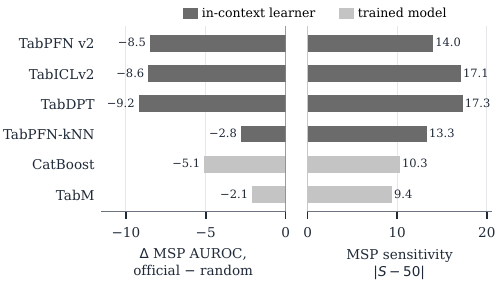}
  \caption{Entity-over-class bias per backbone under the controlled protocol: change of the MSP AUROC from the random to the official
  ID split, and covariate-shift sensitivity $|S-50|$ of the MSP.}
  \label{fig:backbones}
\end{figure}

\subsection{RQ1: Random-Split versus Open-Entity Evaluation}
\label{sec:rq-eval}

\begin{table}[tbp]
  \centering
  \caption{\method{} on three in-context backbones, pooled trials 35--44 over 14 tables. First row: mean AUROC (\%) of
  \method{}; other rows: \method{} minus the detector in mean AUROC, all positive. $^{\ast}$, $^{\ast\ast}$: Holm-corrected
  Wilcoxon $p<0.05$, $p<0.01$ within the backbone.}
  \label{fig:app-backbones}
  \small
  \setlength{\tabcolsep}{5pt}
\begin{tabular}{lccc}
\toprule[1pt]
\textbf{Detector} & \textbf{TabPFN v2} & \textbf{TabICLv2} & \textbf{TabDPT} \\
\midrule
\textbf{\method{} AUROC} & \textbf{81.4} & \textbf{80.8} & \textbf{80.5} \\
\midrule
MSP & $+25.1$$^{\ast}$ & $+25.1$$^{\ast}$ & $+25.0$ \\
Entropy & $+25.0$$^{\ast}$ & $+25.4$$^{\ast}$ & $+25.5$ \\
kNN$^\dagger$ & $+6.7$$^{\ast}$ & $+6.0$$^{\ast}$ & $+5.8$$^{\ast}$ \\
kNN & $+6.9$$^{\ast}$ & $+6.3$$^{\ast}$ & $+6.1$ \\
Maha (x) & $+10.1$$^{\ast}$ & $+9.4$$^{\ast}$ & $+9.2$ \\
Maha (emb.) & $+8.7$ & $+12.0$ & $+8.8$ \\
Emb-kNN & $+14.3$ & $+14.0$ & $+11.5$ \\
IForest & $+10.7$$^{\ast}$ & $+10.0$$^{\ast}$ & $+9.8$ \\
LOF & $+14.3$$^{\ast\ast}$ & $+13.7$$^{\ast}$ & $+13.5$$^{\ast}$ \\
OC-SVM & $+11.4$$^{\ast\ast}$ & $+10.7$$^{\ast\ast}$ & $+10.5$$^{\ast}$ \\
ECOD & $+18.1$$^{\ast\ast}$ & $+17.5$$^{\ast\ast}$ & $+17.2$$^{\ast\ast}$ \\
FoMo-0D & $+14.2$$^{\ast}$ & $+13.5$$^{\ast}$ & $+13.3$$^{\ast}$ \\
OutFormer & $+13.1$$^{\ast\ast}$ & $+12.4$$^{\ast\ast}$ & $+12.2$$^{\ast\ast}$ \\
TabPFN-unsup & $+6.0$ & $+5.3$ & $+5.1$ \\
\methodv{} & $+6.0$$^{\ast\ast}$ & $+5.9$$^{\ast}$ & $+5.7$$^{\ast}$ \\
\bottomrule[1pt]
\end{tabular}
\end{table}

Random-split evaluation overestimates detectors and reorders them. On the six BeyondArena tasks of
\cref{sec:setup}, drawing ID test rows from familiar groups (P1) raises the AUROC of all eight detectors on TabPFN
by 5.2 (IForest) to 15.8 points (embedding Mahalanobis) and changes their ranking. The other two pitfalls of
\cref{sec:pitfalls} distort other aspects of the evaluation. Drawing OOD rows from familiar groups (P2) exaggerates the confidence
defect, since the ID--OOD confidence gap of TabPFN shrinks by 4.0 points with OOD rows from all groups but by 1.7
with OOD rows from new groups. Ranking scores within the test batch (P3) hides the sensitivity to new groups and
lowers $|S-50|$ of all eight detectors to 0.5--3.9 points (kNN: 14.1 to 2.1). 

\begin{figure}[tbp]
  \centering
  \begin{minipage}[t]{0.52\linewidth}
  \centering
  \includegraphics[width=\linewidth]{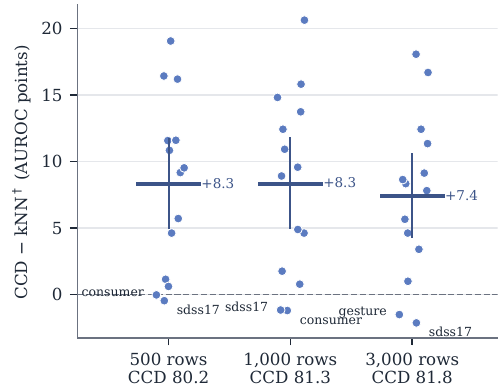}
  \caption{\method{} minus kNN$^\dagger$ in AUROC per table at three context sizes (TabPFN v2, held-out trials 35--39);
  bars show the mean with a 95\% bootstrap interval.}
  \label{fig:app-context}
  \end{minipage}\hfill
  \begin{minipage}[t]{0.44\linewidth}
  \centering
  \includegraphics[width=\linewidth]{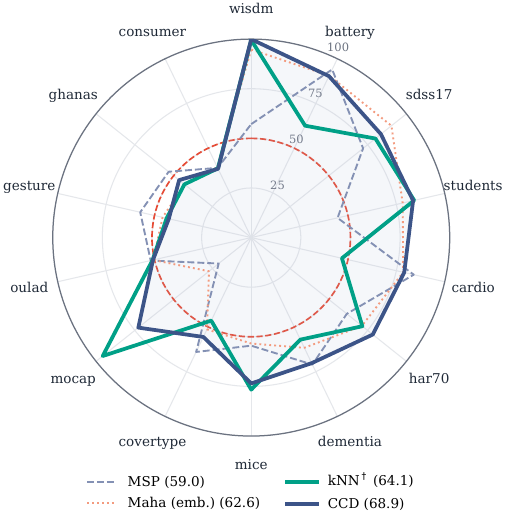}
  \caption{Second-smallest class as OOD (TabPFN v2, held-out trials 35--39): mean AUROC per table; the dashed circle is
  chance, and the legend gives the mean over tables.}
  \label{fig:app-oodclass}
  \end{minipage}
\end{figure}

\subsection{RQ2: Diagnosis of Existing Detectors}
\label{sec:failure}

\textbf{New entities look unknown.} With one context and one calibration per trial, 13 of 14 detectors score ID rows
of new groups as more OOD than held-out ID rows of familiar groups, and their AUROC against the same OOD rows falls by
up to 20.3 points, as \cref{fig:app-fixedctx} shows. Under the controlled protocol, which redraws only the OOD test rows from new groups, the ID confidence of TabPFN v2 \citep{hollmann2025tabpfn}, TabICLv2 \citep{qu2026tabiclv2} and TabDPT \citep{ma2025tabdpt} falls by 5.6, 4.9 and 5.1 points on new groups, and their MSP AUROC by 8.5, 8.6 and 9.2 points,
against 5.1 for CatBoost and 2.1 for TabM, as \cref{fig:backbones} shows. The MSP AUROC drop is largest
on mice (37.7 points) and covertype (10.2).

\textbf{Mechanism.} On cardiotocography, the TFM confidence on correctly classified ID rows falls with distance to the context more steeply than for trees or RealMLP ($-$0.21 for TabPFN versus at most $-$0.09). This decline is also steeper for the TFM on ghanas and mice. These declines are consistent with the confidence encoding familiarity to the context, which a new entity lacks. The trade-off involves a different failure, high confidence on unknown-class rows. On consumer, covertype and dementia the
relation reverses, and rows far from the context receive \emph{higher} confidence. The output scores therefore fall below chance
(TabPFN MSP AUROC 13--23), while a feature-space kNN reaches 83--86 on consumer and dementia. Label-free detectors share the entity-over-class bias but not this failure: they treat a new entity as an outlier. With the detector fixed, kNN and LOF lose 7.3 and 10.0 points of AUROC from familiar to new ID rows and the two embedding scores 15.7
and 20.3. The two families thus fail in different places, which \method{} exploits by combining them.

\subsection{RQ3: Comparison with Existing Detectors}
\label{sec:rq1}

\method{} is the strongest detector overall. Among the 17 detectors in \cref{tab:main}, it has the highest mean AUROC
(81.8) and AUPR (52.4), the lowest FPR95 (50.3) and the best average rank in all three metrics. It gains $+7.4$ points
of AUROC over kNN$^\dagger$ (12/14 tables, $p=0.0012$), and the gain is significant after Holm correction against 14
of the 16 baselines, as the last column of \cref{tab:main} shows. It again has
the highest mean AUROC in the resampling check (81.1) and with TabICLv2 and TabDPT as backbones (81.2 and 81.0, with
gains of $+6.8$ and $+6.6$ over kNN$^\dagger$, $p=0.0017$ and $0.0023$), as \cref{fig:app-backbones} lists. Its gain over kNN$^\dagger$ also persists with at most 500 or 1{,}000 context rows by $+8.3$ points, as \cref{fig:app-context} shows. The gain also persists when every class with at least 30 rows is held out once as OOD. Across the resulting 69 table--class pairs, \method{} gains $+5.8$ points over kNN$^\dagger$ ($p=0.0010$, and $+5.4$ when the classes of each table are averaged first, $p=0.0042$). With the second-smallest class as OOD, \method{} again has the highest mean AUROC among the 11 evaluated detectors at 68.9 against 64.1 for kNN$^\dagger$, as \cref{fig:app-oodclass} shows.

\textbf{Robustness across tables.}\label{sec:rq2} \method{} holds up where the confidence fails:
on the four tables where the confidence is blind (consumer, covertype, dementia and gesture) the output scores fall to 13--24 (MSP) while \method{}
reaches 77.4--87.4.

\begin{table}[tbp]
  \centering
  \caption{Ablation on the held-out trials: mean AUROC (\%) and difference to the full method over the 14 tables.}
  \label{tab:ablation}
  \small
  \setlength{\tabcolsep}{4pt}
\begin{tabular}{lcc}
\toprule[1pt]
\textbf{Variant} & \textbf{AUROC} & $\boldsymbol{\Delta}$ \\
\midrule
\textbf{\method{} (full)} & \textbf{81.8} & -- \\
uniform distance weights & 81.1 & -0.7 \\
w/o label-subspace distance & 79.4 & -2.4 \\
w/o feature-space kNN & 81.7 & -0.0 \\
w/o embedding distance & 80.4 & -1.4 \\
w/o label branch & 78.4 & -3.3 \\
agreement weight (convex fusion) & 79.4 & -2.4 \\
$\theta=0.8$ & 81.1 & -0.7 \\
$\theta=0.95$ & 81.7 & -0.1 \\
\bottomrule[1pt]
\end{tabular}
\end{table}

\subsection{RQ4: Design Analysis}
\label{sec:rq4}

\textbf{Ablation.} The ablation in \cref{tab:ablation} shows that both branches and the fusion rule contribute to \method{}.
Removing the
label branch lowers the mean AUROC by 3.3 points, the label-subspace distance by 2.4 and the embedding distance by 1.4. Convex fusion of the confidence as in \methodv{} costs 2.4
points in line with \cref{thm:fusion}(iii), and the constant $\theta$ is not critical. The contribution of each branch does not depend on the TFM that produces the confidence and the embedding,
because on the other two backbones removing the label-subspace distance costs 2.4--3.1 points and removing the
label branch 2.5--3.3 points.

\textbf{Testable predictions.}\label{sec:theorycheck} Three predictions of \cref{sec:theory} hold on the development trials. The swap estimates match their targets, as \cref{prop:swap} predicts: the Spearman correlation over tables
between $\hat S$ and $S$ is 0.86--0.92. The lower bound on the confidence AUROC of \cref{sec:theory} holds in all 77 table--trial cells of the development trials, and the sharpness
$a$ correlates 0.78 with the AUROC of MSP. The label-subspace distance is less sensitive than the feature-space kNN
($|S-50|$ 12.9 versus 15.3), as the invariance of $d_2$ in \cref{sec:theory} predicts.

\section{Conclusion}
\label{sec:conclusion}

Open-entity OOD detection asks a detector to accept the rows of new entities whose class is known and to reject the
rows of an unknown class. For an in-context TFM both can look equally unfamiliar, because every score it produces is a function of the context. This paper turns that observation into an evaluation and a method. NEO-Tab measures detection on new
entities directly, and on it random-split evaluation overestimates every detector tested and misranks them. Nearly
all detectors also score the known-class rows of new entities as more OOD than those of familiar entities. CCD holds
groups out of the context to obtain, from ID data alone, how known-class rows of new entities score. It calibrates and
weights three distances and the confidence against that reference before fusing them, as the entity--class trade-off requires. CCD has the highest mean AUROC among 17
detectors on 14 grouped tables and gains 7.4 AUROC points over kNN$^\dagger$, a lead that holds on three backbones.
Its three testable predictions hold on the development trials. For a Web service whose rows come from new users or
merchants, this means an OOD detector that is evaluated under the entity shift it will meet in deployment and runs on a frozen TFM without requiring outliers or retraining.

\bibliographystyle{ACM-Reference-Format}
\bibliography{refs}

\end{document}